\documentclass[journal]{IEEEtran}
\usepackage{xcolor,soul,framed} 
\colorlet{shadecolor}{yellow}
\usepackage[pdftex]{graphicx}
\usepackage{caption}
\usepackage{subcaption}
\graphicspath{{../pdf/}{../jpeg/}}
\DeclareGraphicsExtensions{.pdf,.jpeg,.png}
\usepackage[cmex10]{amsmath}
\usepackage{amssymb}
\usepackage{amsthm}
\usepackage{amsmath}
\usepackage{array}
\usepackage{mdwmath}
\usepackage{eqparbox}
\usepackage{url}
\usepackage{tabularx}
\usepackage{booktabs}
\usepackage{algorithm}
\usepackage{algorithmic}
\usepackage{inputenc}
\usepackage{graphicx}

\newtheorem*{remark}{Remark}
\begin{document}
\bstctlcite{IEEEexample:BSTcontrol}



    \title{Quantum Search-based Distributed Beamforming and Re-forming for Collaborative UAS in nextG Communications}

    \title{Adaptive Quantum Dance of Drones: Strategies for Distributed Beamforming and Re-forming for Collaborative UAS Communication for NextG IoT}

    \title{Quantum-Assisted Adaptive Beamforming in UASs Network: Enhancing Airborne Communication via Collaborative UASs for NextG IoT}
    \author{Sudhanshu~Arya,~\IEEEmembership{Member,~IEEE}~        
     and~Ying~Wang,~\IEEEmembership{Member,~IEEE}

  \thanks{(Corresponding author: Ying Wang.)}
  \thanks{Sudhanshu~Arya and Ying~Wang are with the School of Systems and Enterprises, Stevens Institute of Technology, Hoboken, USA (e-mail: sarya@stevens.edu; ywang6@stevens.edu).}
  }

\markboth{
}{Arya \MakeLowercase{\textit{et al.}}: UAS}

\maketitle

\begin{abstract}
This paper introduces a novel quantum-based method for dynamic beamforming and re-forming in Unmanned Aircraft Systems (UASs), specifically addressing the critical challenges posed by the unavoidable hovering characteristics of UAVs. Hovering creates significant beam path distortions, impacting the reliability and quality of distributed beamforming in airborne networks. To overcome these challenges, our Quantum Search for UAS Beamforming (QSUB) employs quantum superposition, entanglement, and amplitude amplification. It adaptively reconfigures beams, enhancing beam quality and maintaining robust communication links in the face of rapid UAS state changes due to hovering. Furthermore, we propose an optimized framework, Quantum-Position-Locked Loop (Q-P-LL), that is based on the principle of the Nelder-Mead optimization method for adaptive search to reduce prediction error and improve resilience against angle-of-arrival estimation errors, crucial under dynamic hovering conditions. We also demonstrate the scalability of the system performance and computation complexity by comparing various numbers of active UASs. Importantly, QSUB and Q-P-LL can be applied to both classical and quantum computing architectures. Comparative analyses with conventional Maximum Ratio Transmission (MRT) schemes demonstrate the superior performance and scalability of our quantum approaches, marking significant advancements in the next-generation Internet of Things (IoT) applications requiring reliable airborne communication networks.

\end{abstract}


\IEEEpeerreviewmaketitle

\section{Introduction}
\IEEEPARstart{T}{he} Internet of Things (IoT) has seen a sharp rise in its popularity and has been considered as a technology that can provide solutions to various social and economic problems \cite{chettri2019comprehensive}. However, with multiple IoT devices embedded in the vicinity environment, to realize the common objective of communication, the presence of data error and falsified perception makes the IoT environment less reliable \cite{lopez2021massive, bhatia2020quantum}. Among many recent advances in disruptive quantum computation technologies, quantum information-assisted communication systems, and networks enjoy special attention; due to many important and interesting quantum features, such as entanglement, superposition, and quantum parallelism \cite{caleffi2023beyond, garcia2014quantum}. Quantum information and computation have emerged as critical enable and catalysts to empower a range of new paradigms from the perspective of next-generation data communication systems, security, computation, and intelligence \cite{mohanty2023quantum, wang2022quantum}.

On the other hand, with the superiority in maneuverability and cost, the recent swarming unmanned aerial systems (UASs) networks are widely deployed in a variety of IoT scenarios \cite{chen2020distributed}. Unmanned aerial systems (UASs) assisted wireless communications can provide long-distance high data transmission with a large coverage area, thereby making it a hot research topic in the next generation communication \cite{xiao2016enabling, zhao2018beam}. Particularly, simple point-to-point communication links over unlicensed frequency bands are commonly utilized to establish links between ground nodes and UASs and lead to constrained performance \cite{liu2019multi}. While applying UASs in next-generation communications, they act as an aerial base station (BS) to serve user equipment (UE), therefore, UASs in each network need to form a beam to cover the intended UEs. In these scenarios, UASs’ positions are essential for their collaboration \cite{chen2020distributed,sun2022collaborative}. However, due to the mobility and hovering of the UASs, the target coverage area may change frequently due to beam distortion \cite{arya2023distributed}. This limitation motivated many researchers to exploit dynamic beamforming by optimizing the 3D positioning of the UASs in the given distributed network \cite{mozaffari2016efficient, 8907440}. Distributed beamforming involves multiple transmitters coordinating their signal emissions to improve reception at various locations \cite{sun2020improving}. It optimizes the overall communication system by allowing decentralized adjustments to transmit signals effectively, enhancing coverage, and ensuring efficient data transfer to diverse receiving devices \cite{hou2023joint, shi2021novel}.

In the conventional analog beamforming technique, a steering vector is generated as the beamforming vector, which enables a narrow beam pointing towards a specific UE. However, it requires an accurate channel stated information between the UASs and the UE for efficient data transmission, which is a challenging task, since the pilot signals, generally being transmitted without adequate beam gains, may not be effectively detected by the receiver owing to the severe losses due the channel fading and UASs hovering. Furthermore, considering the size and power constraints of a UAS, the existing beamforming techniques may not be efficient for UAS-assisted cellular communications.

To this end, we propose a quantum-based unified 3D beamforming and beam re-forming framework for hovering impaired UAS-assisted wireless networks. Quantum computation and information represent a disruptive technology capable of supporting functionalities that have no direct counterpart in the classical world, such as exponential increase of quantum computing power, secure communications, advanced quantum sensing techniques, and blind computing \cite{rezai2021quantum, chandra2021direct}. Motivated by this recent advancement in quantum computing, we develop a new framework for a UAS-assisted communication network. We call it quantum search for UAS beamforming (QSUB) and show for the first time that by utilizing quantum computation, it is possible to construct a directional communication with sufficient beam gains by dynamically controlling the beam distortion in a UAS-assisted distributed network. QSUB is shown to perform well under random hovering conditions however it is worth mentioning that QSUB is catering to the zero angle-of-arrival (AoA) estimation error and the performance is compromised as the AoA estimation error increases. To overcome this limitation of the QSUB, we further propose an advanced QSUB (Q-P-LL) to account for the AoA estimation error. We show that Q-P-LL outperforms QSUB significantly as the AoA estimation error increases.

\subsection{Contributions}
The main contributions of this paper are listed below.
\begin{itemize}
    \item This work identifies the critical open research challenges and presents a new and evolutionary quantum-inspired positioning and beam-steering method, where principles of quantum computations are utilized for formulating a distributed beam-steering method for hovering impaired UAS-assisted NextG IoT network. We consider the problem of beam distortion and characterized it by positioning and synchronization errors due to the hovering and rotation motions of the UASs, and channel-induced errors, including phase distortion and interference. We present a detailed quantum architecture of the proposed framework. The proposed QSUB is versatile and can be implemented on both quantum and classical computing architectures.
    \item Furthermore, to improve the accuracy of the QSUB over non-zero AoA estimation error, a new optimized framework Q-P-LL, based on the Nelder-Mead numerical optimization, is presented that searches adaptively based on the topography of the fitness landscape. We show that Q-P-LL significantly improves the beamforming performance by reducing the UAS prediction error. It is demonstrated that, compared to the QSUB, Q-P-LL performance is less sensitive to the AoA estimation error. We show that to achieve a success probability of the proposed Q-P-LL greater than one-half for locking to the correct position, the quantum oracle must be called $\Delta(K_u^{N_u})$ times, where $K_u$ is the number of quantum states for individual UASs and $N_u$ is the number of UAS in the network.
\end{itemize}

\begin{figure*}
    \centering
        \includegraphics[width=5.2in, height=3.1in]{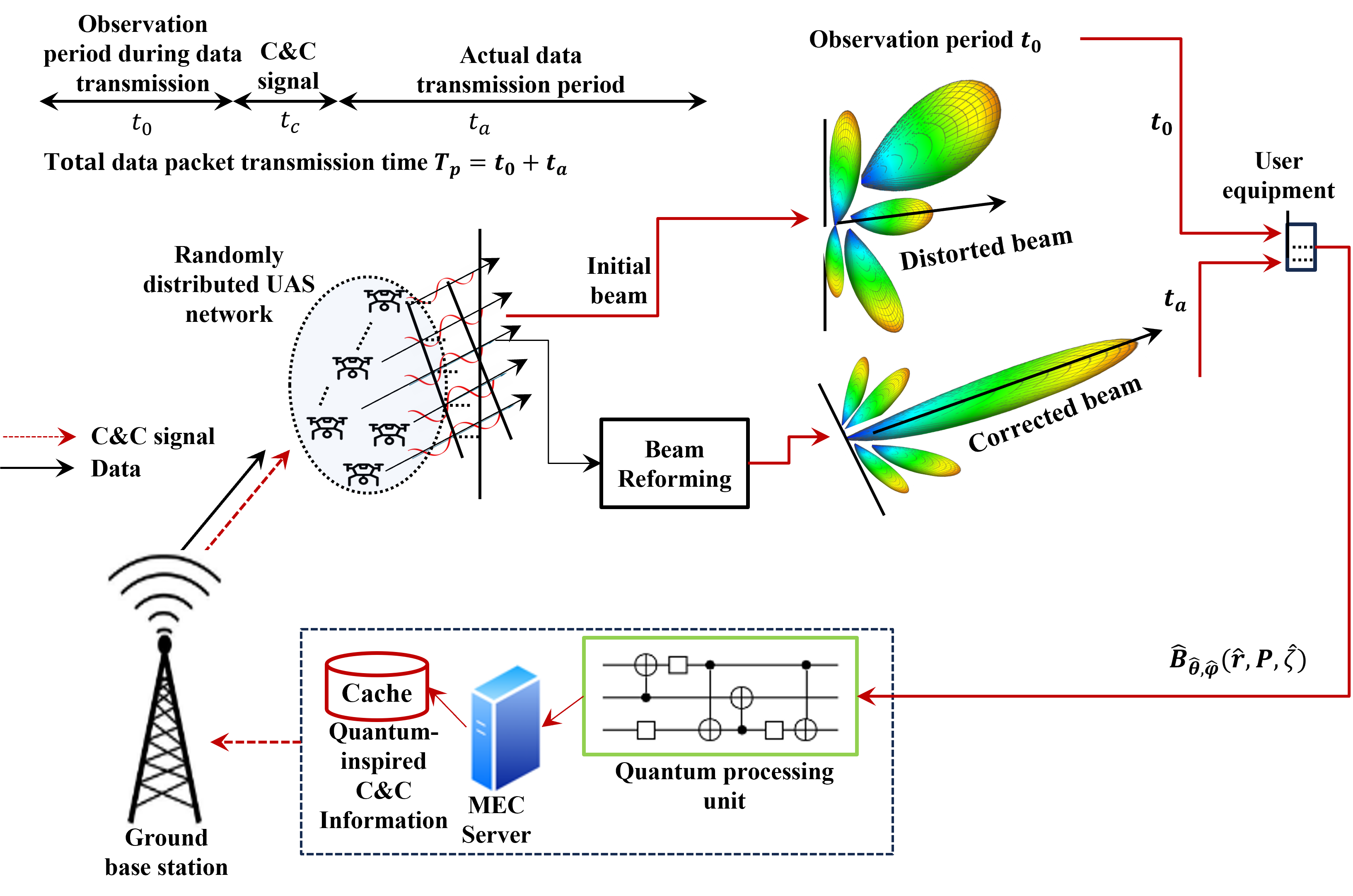}
    \caption{Illustration of the hovering impact and beam distortion.}
    \label{FIG-1}
\end{figure*}
\section{System Model and Problem Formulation}
The system under consideration is illustrated in Fig. \ref{FIG-1}, where the command and control signal is abbreviated as C$\&$C. EC stands for multi-access edge computing and provides computing capabilities and a service environment at the edge of the network. The proposed system model enables computation offloading by the user equipment. The user equipment offloads the task of quantum computation to the MEC server. As depicted in the figure, the quantum computation unit is located inside the MEC unit. To minimize the backhaul delay, the MEC server is located near the ground base station. Let $\mathcal{N}_U$ be the total number of distributed UASs in a network. 
Let $ \mathbf{P_k}(t) \equiv \mathbf{P}(r, \theta, \phi) $ represents a random position of the $k$th UAS at time instant $t$, such that $ \mathbf{P}(r, \theta, \phi) $ $ \triangleq [r\sin\theta\cos\phi, r\sin\theta\sin\phi, r\cos\theta] \in \mathcal{R} \subset \mathbb{R}^C$, where $ \mathbb{R}^C $ represents $ C $-dimensional Euclidean domain. Figure \ref{FIG-1} represents a random three-dimensional hovering impact. $ r = \Vert \mathbf{P} \Vert \in [0,\infty) $ denotes the radius and $ \Vert \cdot \Vert $ represents the Euclidean norm. $ \theta \in [0, \pi] $ denotes the collatitude measured with respect to the positive $ Z- $axis. $ \phi \in [0, 2\pi) $ denotes the longitude measured with respect to the positive $ X- $ axis in the $ XY- $plane. We consider a ball geometry, mathematically represented by $ \mathcal{R}_B \triangleq \{\mathbf{P} \in \mathbb{R}^3 : \Vert\mathbb{R}^3\Vert \leq \mathcal{R}\}$, where $ \mathcal{R} $ denotes the radius of the ball. It ensures that the distance between the $n$th and the $k$th UAVs, be such that, $d_{k,n} \geq \mathcal{R}$, where,  $n, k \in \{1, 2, \cdot\cdot\cdot, N_u\}$.  

The generalized beam-pattern for $N_u \in \mathcal{N}_U$ optimal UASs located at position $\mathbf{r}_k  \buildrel \Delta \over = \left[ {x_k ,y_k ,z_k} \right]^T  \in \mathbb{R}^3, k = \{1, 2, \cdot\cdot\cdot, N_u\}$ and transmitting with powers $P_1, P_2, \cdot\cdot\cdot, P_{N_u}$ and phases $\zeta_1, \zeta_2, \cdot\cdot\cdot, \zeta_{N_u}$ can be expressed as
\begin{equation}
    \begin{split}
& B_{\theta ,\varphi } \left( {\mathbf{r},\mathbf{P},\mathbf{\zeta} } \right) = \left| {\sum\limits_{k = 1}^{N_u} {P_k w_k \exp \left[ {j\left( {\zeta _k  + \frac{{2\pi }}{\lambda }x_k \cos \varphi \sin \theta } \right.} \right.} } \right. \\ 
& \hspace{1.8cm}\left. {\left. {\left. { + \frac{{2\pi }}{\lambda }y_k \sin \varphi \sin \theta  + \frac{{2\pi }}{\lambda }z_k \cos \theta } \right)} \right]} \right| \\
    \end{split}
    \label{EQ-1},
\end{equation}
where $\lambda$ is the carrier wavelength, $\mathbf{r} = \left[\mathbf{r}_1^T, \mathbf{r}_2^T, \cdot\cdot\cdot, \mathbf{r}_{N_u}^T\right] \in \mathbb{R}^{3K}$. $\theta \in [-\pi, \pi]$ and $\varphi \in [-\pi, \pi]$ represent the elevation and the azimuthal angles, respectively.

\begin{figure}
    \centering
        \includegraphics[width=0.9\linewidth]{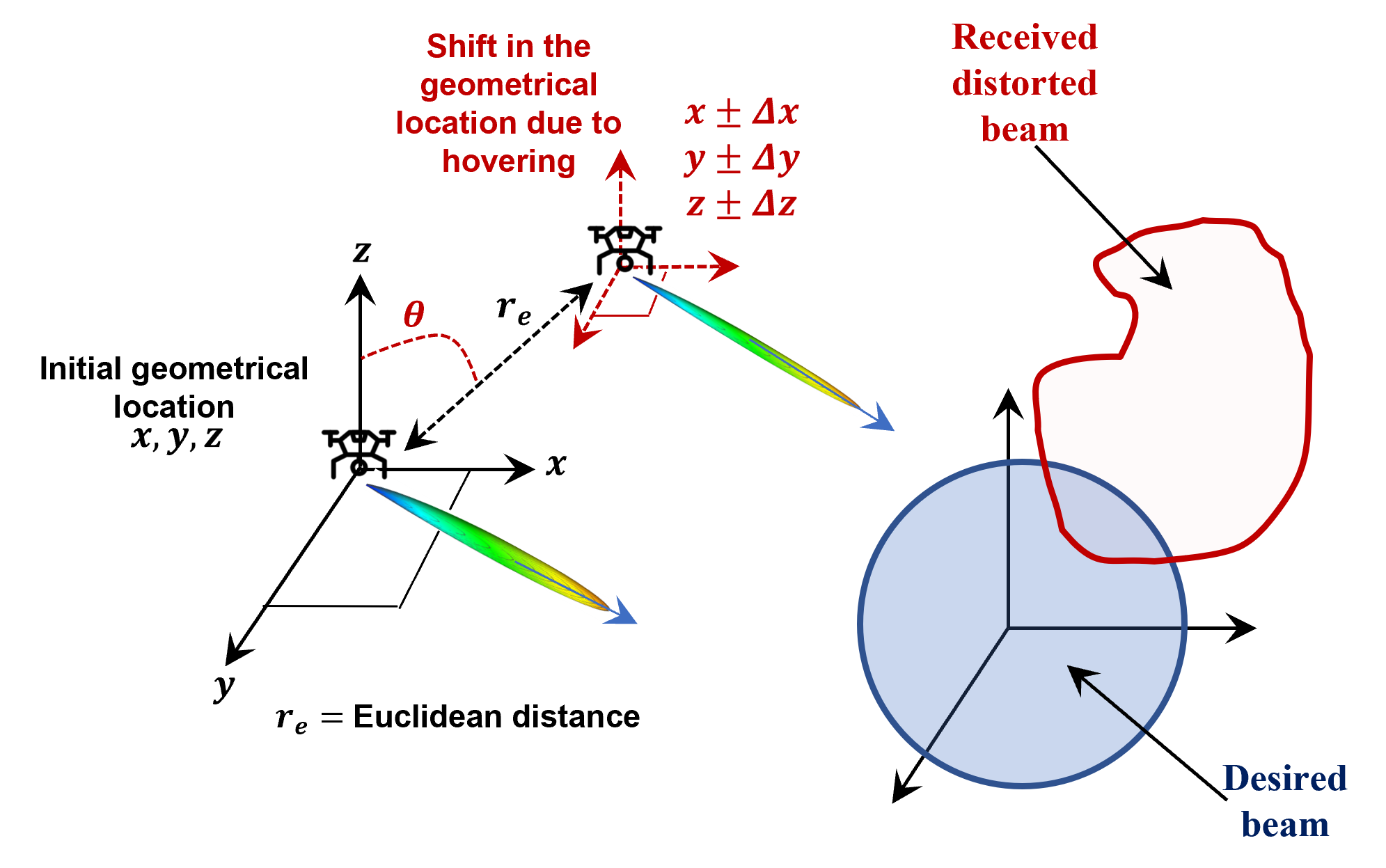}
    \caption{Illustration of the hovering impact and beam distortion.}
    \label{FIG-3}
\end{figure}

As illustrated in Fig. \ref{FIG-1}, the rotational motion of the UAS can be categorized into three types: yaw $\xi$, pitch $\gamma$, and roll $\Theta$. We characterize the beam distortion and misalignment as system-induced errors, which include positioning and synchronization errors due to the hovering and rotational motion, and channel-induced errors, including phase distortion and interference. These errors can be modeled as perturbations in the location and the orientation parameters $\{\Delta x, \Delta y, \Delta z, \Delta \theta, \Delta \varphi\}$, and the phase parameter $\Delta\zeta$. The distorted beam pattern can then be written as
\begin{equation}
    \begin{split}
& \hat B_{\hat \theta ,\hat \varphi } \left( { \mathbf{\hat r},\mathbf{P}, \mathbf{\hat \zeta }} \right) = \left| {\sum\limits_{k = 1}^{N_u} {P_k w_k \exp \left[ {j\left( {\hat \zeta _k  + \frac{{2\pi }}{\lambda }\hat x_k \cos \hat \varphi \sin \hat \theta } \right.} \right.} } \right. \\ 
& \hspace{1.8cm} \left. {\left. {\left. { + \frac{{2\pi }}{\lambda }\hat y_k \sin \hat \varphi \sin \hat \theta  + \frac{{2\pi }}{\lambda }\hat z_k \cos \hat \theta } \right)} \right]} \right| \\ 
    \end{split}
    \label{EQ-2},
\end{equation}
where ${\hat \zeta } = \zeta \pm \Delta\zeta$, ${\hat{x}_k} = x_k \pm \Delta x_k$, ${\hat{y}_k} = y_k \pm \Delta y_k$, ${\hat{z}_k} = z_k \pm \Delta z_k$, ${\hat{z}_k} = z_k \pm \Delta z_k$, ${\hat{\theta}_k} = \theta_k \pm \Delta \theta_k$, and ${\hat{\varphi}_k} = \varphi_k \pm \Delta \varphi_k$. Figure \ref{FIG-3} illustrates the hovering impact on the beamforming.

\begin{figure}
    \centering
        \includegraphics[height=1.5in]{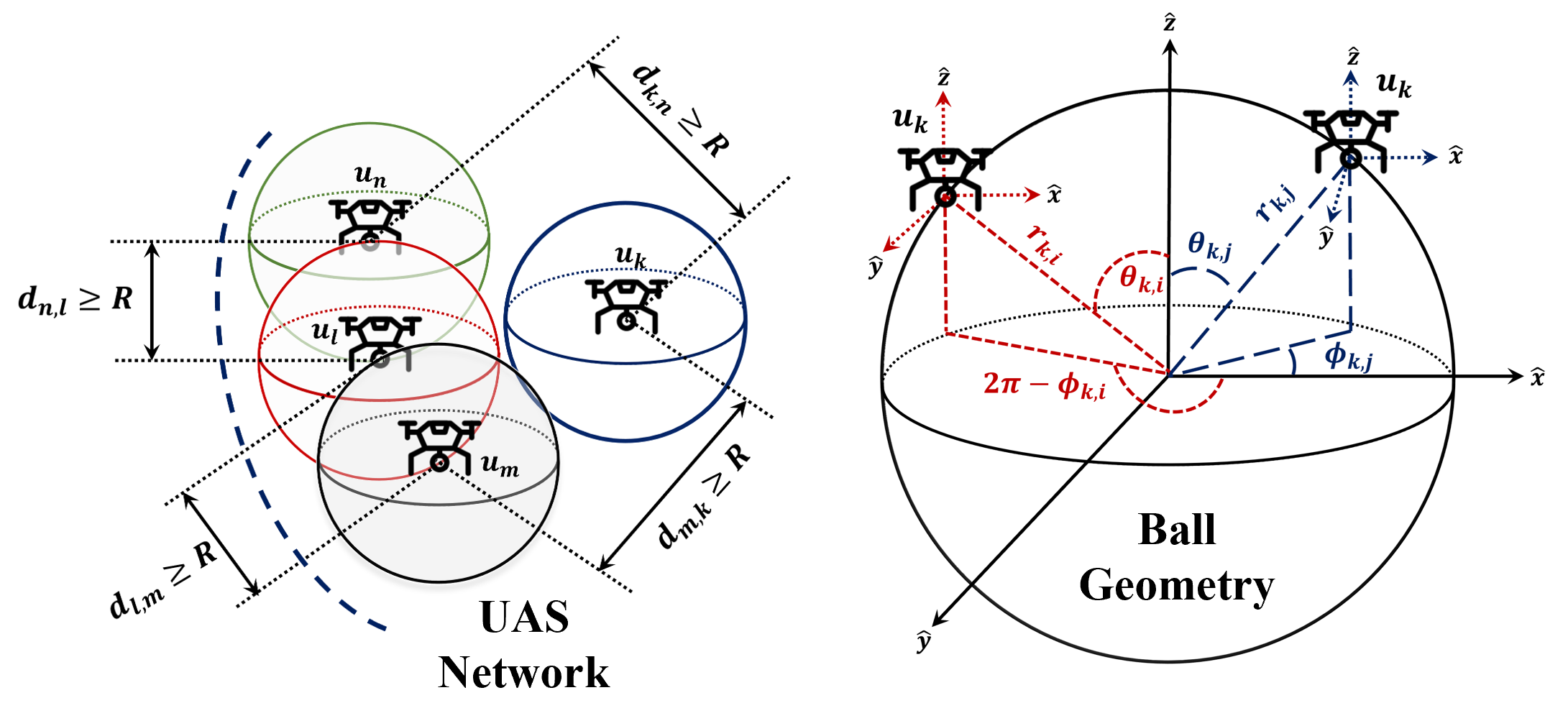}
    \caption{UASs arrangement in a ball geometry.}
    \label{FIG-2}
\end{figure}

We construct an optimization problem where distributed beamforming where each UAS adjusts its coordinates $x_k, y_k, z_k$, rotational angles $\xi, \gamma, \Theta$, and phase $\zeta_k$ for $k \in \{1, 2, \cdot\cdot\cdot, N_u\}$. With this aim, the optimization problem can be written as the minimization of the following objective function:
\begin{equation}
    \begin{split}
 & \mathop {\min }\limits_{\mathbf{r}, \zeta } J\left( {\mathbf{r},\mathbf{P},\zeta } \right) =  \\ 
 & \frac{1}{4}\int\limits_{ - \pi }^\pi  {\int\limits_{ - \pi }^\pi  {\left\| {B_{\theta ,\varphi } \left( {\mathbf{r},\mathbf{P},\mathbf{\zeta} } \right) - \hat B_{\hat \theta ,\hat \varphi } \left( { \mathbf{\hat r},\mathbf{P}, \mathbf{\hat \zeta }} \right)} \right\|_2^2 d\theta d\varphi } }  \\
    \end{split}
    \label{EQ-3}.
\end{equation}

\begin{remark}
    We like to point out that the term beam re-forming in our work involves dynamically detecting and adapting to the changing positions and orientations of UASs. It is defined as reshaping the antenna's distorted beam pattern and minimizing the side lobes, which is impaired due to the conditions of UAVs hovering and moving. As illustrated in Fig. \ref{FIG-1}, the quantum-inspired C\&C signals are generated by the quantum processing unit based on the principle of the proposed quantum computation algorithm QSUB or Q-P-LL. To differentiate the coined term beam re-forming from the most commonly used terms beam tracking and beam steering, we like to emphasize that the proposed quantum-inspired beam re-forming technique primarily focuses on correcting distortion caused by UASs hovering through real-time adjustments, ensuring a stable and accurate beam at the receiver through the application of quantum computation. In contrast, the term beam tracking {\rm \cite{9547829}}, emphasizes continuous adaption to changing conditions based on the feedback from the tracking system, such as a moving target or varying channel characteristics. Whereas, the term beam steering is a broader concept that includes the ability to adjust the beam direction by adjusting the phase and/or amplitude of signals at each fixed-location antenna element {\rm \cite{10161729}}.
\end{remark}

The optimization problem in (\ref{EQ-3}) can be cast in a black-box query framework. For example, given a function $f:\{\mathbf{\hat r}, \mathbf{\hat \zeta } \} \rightarrow \{0, 1\}$ on some finite set $D = \{\mathbf{\hat r}, \mathbf{\hat \zeta }\}$, find $\hat{r} \in \mathbf{\hat r}$ and $\hat{\zeta} \in \mathbf{\hat \zeta }$ such that $f(\hat{r}, \mathbf{\hat{\zeta}}) = 1$. The hardest instance of this problem is when $f(\hat{r}, \mathbf{\hat{\zeta}}) = 1$ only for a unique $(\hat{r}, \mathbf{\hat{\zeta}})$. If $\mathcal{S}$ is the size of the search problem, we show in Section \ref{SECTION-III}, that the proposed quantum-inspired Q-P-LL can solve this problem using $\mathcal{O}(\sqrt{\mathcal{S}})$ superposition queries, whereas classically solving this requires $\mathcal{O}(\mathcal{S})$ queries in the worst case and $\mathcal{O}(\mathcal{S})/2$ queries on average.

The proposed QSUB framework is a quantum routine that accomplishes two tasks:
\begin{itemize}
    \item In the first stage, it identifies the best UAS links $N_u \leq \mathcal{N}_U$ in the given distributed network to obtain the beamforming.
    \item During the second stage, considering the random hovering distribution, it predicts the precise locations of the active UAS links utilized in the beamforming and feedback to the control unit for beam re-forming.
\end{itemize}

\subsection{{QSUB}}
\begin{figure*}
    \centering
\includegraphics[width=14.8 cm, height=4.7cm]{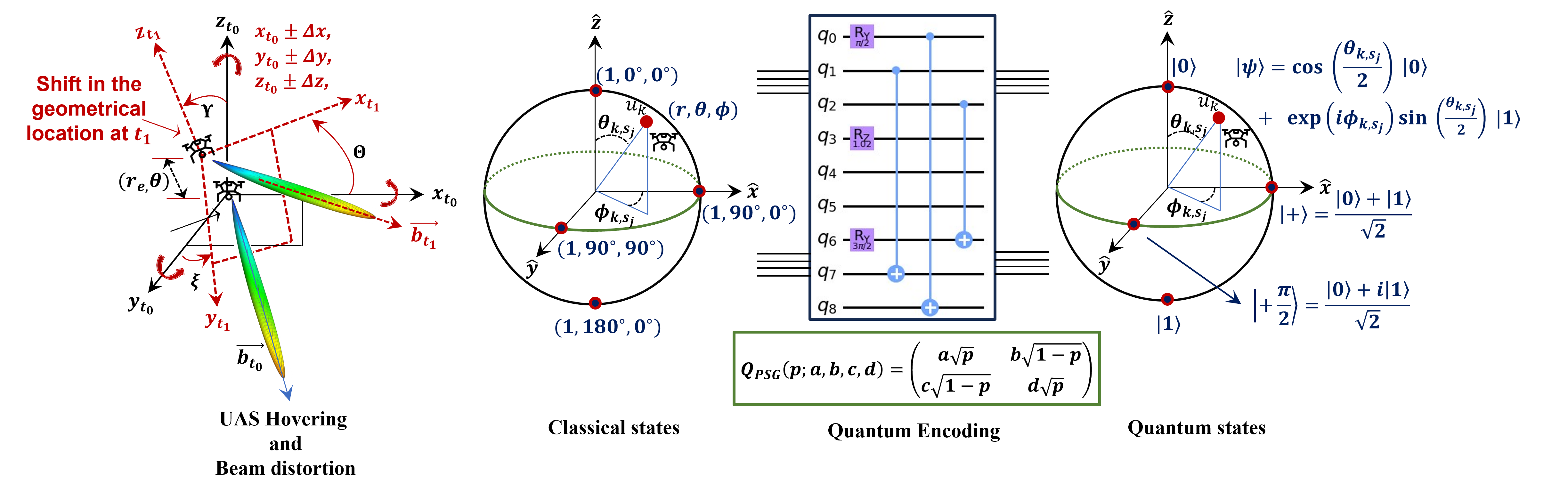}
\caption{Encoding classical information into quantum states using amplitude encoding and entanglement. In this mapping of classical to the quantum state, the quantum circuit is constructed assuming $ r = \Vert \mathbf{P} \Vert = 1 $ with hovering resolution $r_r = 0.25$ m whereas $ \theta \in [0, \pi] $ and $ \phi \in [0, 2\pi) $ have hovering resolutions, $r_\theta$ and $r_\phi$, of $45$ degrees.}
\label{FIG-4}
\end{figure*}

Let $\mathbf{E}_n, n = \{1, 2, \cdot\cdot\cdot, N_u\}, \mathbf{E}_n \in \mathbf{E}$, is the three-dimensional space corresponds to the $n$th UAS that ensembles all the possible random states of $n$th UAS. We first convert the classical search space to the quantum search space. The first stage of the QSUB consists of labeling the states encoding the solution of the search problem and then amplifying their measurement probability. It involves encoding the classical information into quantum states using amplitude encoding and entanglement, as illustrated in Fig. \ref{FIG-4}. As depicted in Fig. \ref{FIG-4}, To encapsulate all the possible quantum states on the Bloch sphere, we present a generalized gate and call it a quantum position synchronization gate $(Q_{PSG})$.
{\begin{equation}
{Q_{PSG}}\left( {p;a,b,c,d} \right) = \left( {\begin{array}{*{20}{c}}
   {a\sqrt p } & {b\sqrt {1 - p} }  \\
   {c\sqrt {1 - p} } & {d\sqrt p }  \\
\end{array}} \right)
\end{equation}}
For $Q_{PSG}$ to be the quantum gate, it must be unitary, that is, $Q_{PSG}Q_{PSG}^\dag = I$, where $I$ is the identity matrix \cite{lai2022learning}. It can be readily shown that for $p = 0.5, a = 1, b = 1, c = 1$, and $d = -1$, $Q_{PSG}$ becomes quantum Hadamard gate $H_d$. $H_d$ is utilized in the quantum circuit in QSUB and Q-P-LL to transform the quantum states. The quantum encoding is performed through a quantum circuit consisting of quantum gates. To encode the classical information on the quantum states, we consider rotation operators $R_y$ and $R_z$, and controlled-NOT (CNOT) quantum gates \cite{nielsen2010quantum}. The $R_y$ gate is a single-qubit rotation through any angle $\phi$ (radians) around the $y-$axis. Mathematically, $R_y$ can be expressed as
{\begin{equation}{R_y}\left( \phi  \right) = \left( {\begin{array}{*{20}{c}}
   {\cos \frac{\phi }{2}} & { - \sin \frac{\phi }{2}}  \\
   {\sin \frac{\phi }{2}} & {\cos \frac{\phi }{2}}  \\
\end{array}} \right).
\end{equation}}
Similary, $R_z$ represents the single-qubit rotation around the $z-$axis through any angle $\phi$ and can be expressed as 
{\begin{equation}{R_z}\left( \phi  \right) = \left( {\begin{array}{*{20}{c}}
   {\exp \left( { - j\frac{\phi }{2}} \right)} & 0  \\
   0 & {\exp \left( {j\frac{\phi }{2}} \right)}  \\
\end{array}} \right).
\end{equation}}

\begin{remark}
    In accordance with the quantum superposition principle, as depicted in Fig. \ref{FIG-4}, similar to the classical computation, the number of quantum states is described by the list of $K_u^{N_u}$ amplitudes, one for each possible length-$\log_2(K_u^{N_u})$ bit string {\rm\cite{jordan2018quantum}}.
\end{remark}

Define a \textit{unitary} matrix $U$ such that $U^{\dag}U = I$. Let, at time instant $t_0$, the UAS system under test be in any random initial quantum state $\left| s_j\rangle \right., j = \{1, 2, \cdot\cdot\cdot, K_u^{N_u}\}$, with respective probability ${\rm P}_{r,j}$. The ensemble of pure quantum states can be defined as $\{{\rm P}_{r,j}, \left| s_j\rangle \right.\}$.

QSUB search through a space of $K_u^{N_u}$ quantum amplitudes in the quantum search space $\mathcal{S}$. Let the quantum search problem has exactly $\mathcal{S}_M \in \mathcal{S}_{\rm opt}$ solutions, such that, $\mathcal{S}_{\rm opt} = \{\mathcal{S}_{\rm opt}^{(1)}, \mathcal{S}_{\rm opt}^{(2)}, \cdot\cdot\cdot, \mathcal{S}_{\rm opt}^{(\mathcal{S}_M)}\} \in \mathcal{S}; 1 \leq \mathcal{S}_M \leq K_u^{N_u}$. As illustrated in Fig. \ref{FIG-4}, the $K_u^{N_u}$ number of quantum amplitudes are stored in $\log_2(K_u^{N_u})$ qubits. Therefore, any arbitrary instance of the search problem can be expressed by a function $f$, which corresponds to the quantum state $s_i \in \mathcal{S}$, takes an integer $i$, in the range $0$ to $K_u^{N_u} - 1$, such that, $f(i) = 1$ if $s_i \in \mathcal{S}_{\rm opt}$ is the optimal solution and $f(i) = 0$ if $s_i$ is not the optimal solution to the quantum search problem.

\begin{figure*}[t]
    \centering
        \includegraphics[width=15.5 cm, height=8.0 cm]{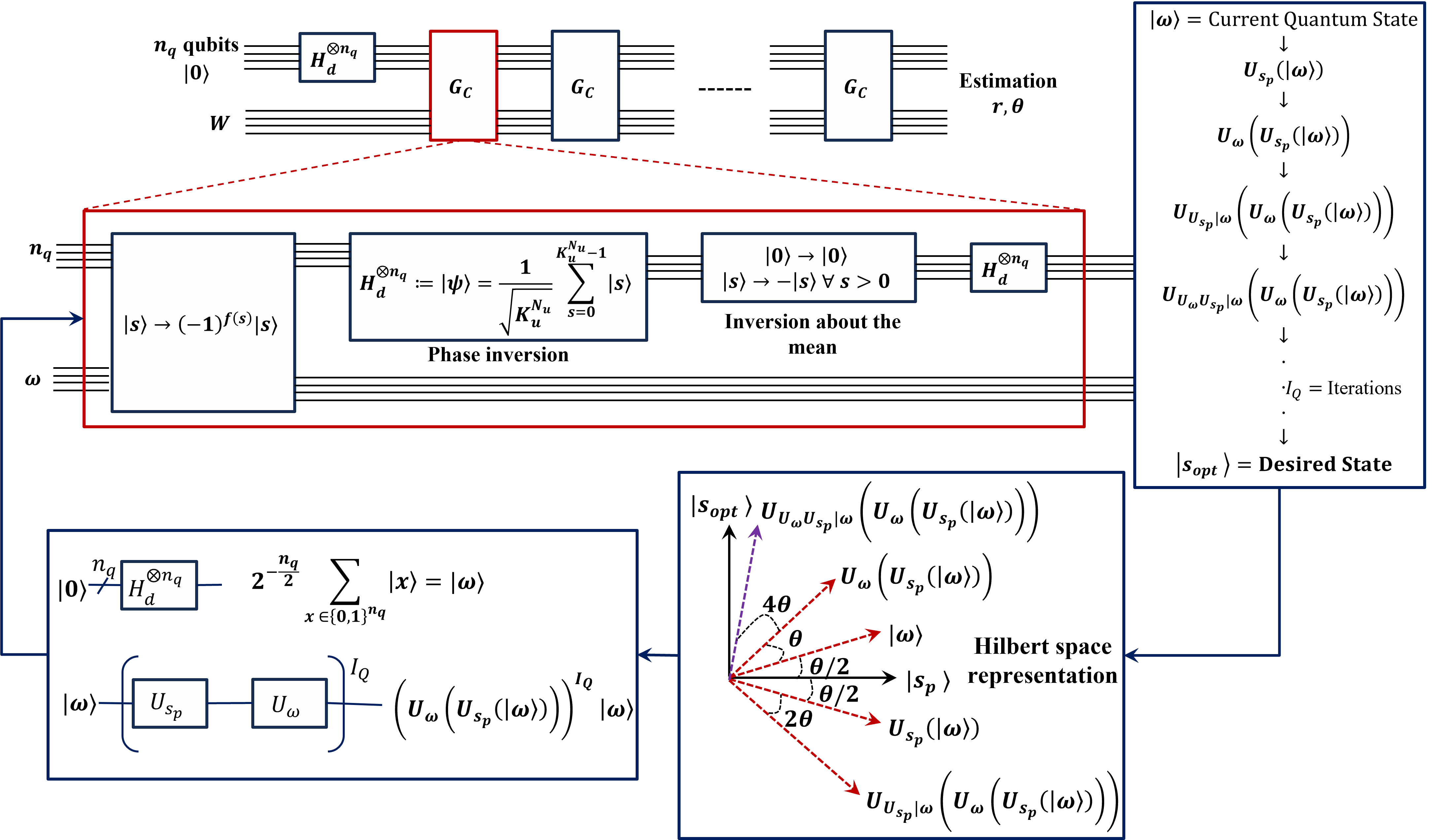}
    \caption{QSUB routine: Quantum circuit to implement the search algorithm for optimal UASs placement prediction.}
    \label{FIG-6}
\end{figure*}

Intuitively, the quantum circuit represented in Fig. \ref{FIG-6}, is the unitary operation defined as
{\begin{equation}
\left| {s_i } \right.\rangle \left| q \right.\rangle \rightarrow \left| {s_i } \right.\rangle \left| {q \oplus f\left( {s_i } \right)} \right.\rangle
\end{equation}}
where $\left| {s_i } \right.\rangle$ represents the index register, $\left| q \right.\rangle$ denotes the single qubit which flips if $f(s_i) = 1$, and remains unchanged otherwise.
\begin{remark}
    More precisely, the quantum circuit presented in {\rm Fig.} \ref{FIG-6}, can determine the optimal quantum state $\mathcal{S}_{\rm opt}$ corresponds to the optimal classical state $\mathbf{E}_i$ by checking whether $s_i$ is a solution to the optimal beamforming problem by preparing $\left| {s_i } \right.\rangle \left| 0 \right.\rangle $, applying to the circuit $\mathcal{O}$, and then determining the output of $\mathcal{O}$ to see if the qubit $q$ has been flipped to quantum state $\left| 1\rangle \right.$ 
\end{remark}

We define the density operator as
\begin{equation}
\rho _Q  = \sum\limits_{j = 1}^{K_u^{N_u } } {\left. {{\rm P}_{r,j} } \right|s_j \rangle \langle \left. {s_j } \right|}
\end{equation}
After the evolution occurred, at time $t_1$, the UAS system will be in the random state $U\left| {s_j \rangle } \right.$ with probability ${\rm P}_{r,j}$. Therefore, the evolution of the density operator $\rho _Q$ can readily be described as \cite{clouatre2022linear}
\begin{equation}
\rho _Q  = \sum\limits_{j = 1}^{K_u^{N_u } } {\left. {P_{r,j} } \right|s_j \rangle \langle \left. {s_j } \right|} \mathop  \to \limits^U \sum\limits_{j = 1}^{K_u^{N_u } } {\left. {{\rm P}_{r,j} U} \right|s_j \rangle \langle \left. {s_j } \right|U^\dag  }  = U\rho _Q U^\dag
\end{equation}

The quantum states obtained after encoding classical state space using quantum amplitude encoding and entanglement, are applied to the quantum circuit, as illustrated in Fig. \ref{FIG-6}, to identify the optimal solution $\mathcal{S}_{\rm opt}$.
The conceptual circuit implementation of the QSUB is depicted in Fig. \ref{FIG-6}. Inspired by Grover's algorithm \cite{755839}, and as illustrated in Fig. \ref{FIG-6}, the QSUB involves the operation of inversion about the mean. $G_C$ denotes the quantum circuit inspired by Grover's algorithm. In Fig. \ref{FIG-6}, $H_d$ is the Hadamard transformation matrix is defined as
\begin{equation}
H_d  = \frac{1}{{\sqrt 2 }}\begin{array}{*{20}c}
   {\begin{array}{*{20}c}
   {{\rm  \mathbf{0} }} \hspace{1cm}& {\rm \mathbf{1}}  \\
\end{array}}  \\
   {\begin{array}{*{20}c}
   {\rm \mathbf{0}}  \\
   {\rm \mathbf{1}}  \\
\end{array}\left[ {\begin{array}{*{20}c}
   {\left( { - 1} \right)^{0 \wedge 0} } & {\left( { - 1} \right)^{0 \wedge 1} }  \\
   {\left( { - 1} \right)^{1 \wedge 0} } & {\left( { - 1} \right)^{1 \wedge 1} }  \\
\end{array}} \right]}  \\
\end{array}
\label{EQ-1}
\end{equation}
Applying a Hadamard transform to all the qubits representing quantum states, followed by a conditional phase shift, increases the amplitude of the optimal quantum state and decreases the amplitude of the others.

Following Eq. (\ref{EQ-1}), with $K_u^{N_u}$ number of total quantum states, the Hadamard matrix for the system under consideration can be expressed as

\begin{equation}
\begin{split}
   & Q\left( {\left| 0 \right.\rangle ^{ \otimes \log _2 \left( {K_u^{N_u } } \right)} } \right) = \left| s \right.\rangle  = \alpha \left| {s_A \rangle  + } \right.\left. \beta  \right|s_{\frac{1}{A}} \rangle  \\ 
 & : = \frac{1}{{\sqrt {2^{\log _2 \left( {K_u^{N_u } } \right)} } }}\sum\limits_{s = 0}^{\log _2 \left( {K_u^{N_u } } \right) - 1} {\left| s \right.\rangle \left( {\frac{{|0\rangle  - \left| {1\rangle } \right.}}{{\sqrt 2 }}} \right)}  \\
\end{split}
\label{EQ-2},
\end{equation}
where $|\alpha\rangle$ and $|\beta\rangle$ are the normalized quantum states of the UASs defined as
\begin{equation}
\left| \alpha  \right.\rangle  \equiv \frac{1}{{\sqrt {K_u^{N_u }  - M} }}\sum\limits_{s \notin S_M } {\left. {\left| s \right.} \right\rangle }; \hspace{0.5cm} \left| \beta  \right.\rangle  \equiv \frac{1}{{\sqrt M }}\sum\limits_{s \in S_M } {\left. {\left| s \right.} \right\rangle }
\label{EQ-8}.
\end{equation}
It is to be noted that the state $|\alpha\rangle$ corresponds to the sum over all $s$ which are not the solutions, whereas $|\beta\rangle$ corresponds to the sum over all $s$ which are the solution to our problem of finding the optimal UASs.

We define a matrix $A$ such that
\begin{equation}
A = \left[ {\begin{array}{*{20}c}
   {2^{ - N_u \log _2 K_u } } & {2^{ - N_u \log _2 K_u } } &  \cdots  & {2^{ - N_u \log _2 K_u } }  \\
   {2^{ - N_u \log _2 K_u } } & {2^{ - N_u \log _2 K_u } } &  \cdots  & {2^{ - N_u \log _2 K_u } }  \\
    \vdots  &  \vdots  &  \ddots  &  \vdots   \\
   {2^{ - N_u \log _2 K_u } } & {2^{ - N_u \log _2 K_u } } &  \cdots  & {2^{ - N_u \log _2 K_u } }  \\
\end{array}} \right]
\end{equation}
It can readily be shown that multiplying any quantum state by $A$ will generate a state where each term will be the average of all the terms. Let $\mathcal{S}_{\rm opt} \in \mathcal{S}$ be the optimal state or the solution of the quantum search problem. 
The objective function is:
\begin{equation}
f\left( {\mathbf{E},\mathbf{\Theta}} \right) = \frac{1}{{N_u }}\sum\limits_{j = 1}^{N_u } {\left\| {\mathbf{E}_j  - \mathbf{\Theta} _j } \right\|^2 }
\end{equation}
The term $H_d^{ \otimes \log _2 \left( {K_u^{N_u } } \right)}$ is expressed as
\begin{equation}
    H_d^{ \otimes \log _2 \left( {K_u^{N_u } } \right)}  = \underbrace {H_d  \otimes H_d  \otimes  \cdot  \cdot  \cdot  \otimes H_d }_{\log _2 \left( {K_u^{N_u } } \right){\rm   terms}}
\end{equation}
\begin{table}
	\renewcommand{\arraystretch}{1.3}
	\caption{Notations}
	\label{TABLE-I}
	\centering
	\begin{tabular}{ l p{4cm} } 
		\hline
		\textbf{Parameter} & \textbf{Description} \\
		\hline
		$ \mathcal{N}_U$ & Number of distributed UASs in a network\\
        \hline
		$ N_u \leq \mathcal{N}_U$ & Number of UASs utilized for beamforming\\
		\hline
        $\mathbf{E}$ & Classical search space \\
        \hline
        $\mathcal{S}$ & Quantum search space \\
        \hline
        $ K_u $ & Number of quantum states a UAS can attain; $K_u \in \mathcal{S}$ \\
		\hline
        $G_C$ & Grover iteration circuit \\
        \hline
        $\mathcal{S}_{\rm opt} \in \mathcal{S}$ & Solution space \\
        \hline
        $r_e$ & Euclidean distance (error) \\
        \hline
        $\Theta$ & Target (actual) coordinates \\
        \hline
        $r_r$ & Resolution of $r \in \mathcal{R}$ describing the ball geometry \\
        \hline
        $r_\Omega$ & Resolution of $\phi$ and $\theta$ describing the ball geometry \\
        \hline
        $k_{k,n}$ & Distance between the $k$th and the $n$th UAS \\
        \hline
        $I_{Q}$ & Number of iterations \\
        \hline
    \end{tabular}
\end{table}

\begin{algorithm}
	\renewcommand{\algorithmicrequire}{\textbf{Input:}}
	\renewcommand{\algorithmicensure}{\textbf{Output:}}
	\caption{QSUB}
        \textbf{Input}: Number of UASs in the network: $N_u$ \\
        \textbf{Input}: Number of quantum states of a UAS: $K_u$ \\
        \textbf{Input}: Iterations: $I_{Q}$ \\
        \textbf{Input}: Oracle $\mathcal{O}$ to perform the transformation: $O|s\rangle q\rangle  = |s\rangle q \oplus f\left( s \right)\rangle$ \\
        \textbf{Initialize}: Quantum states:  $\left| {\mathbf{0}\rangle  = } \right.\left| 0 \right.\rangle ^{ \otimes \log_2({K_u^{N_u }}) } \left| {0\rangle } \right.$ \\
        \textbf{Initialize}: Optimal performance:  $F_{Max} \leftarrow - \infty$ \\
        \textbf{Define}: Empty list: Optimal\_Combination $C_{opt}^{N_u} = [\hspace{0.1cm}]$ \\
        \begin{algorithmic}[1]
        \STATE UAV\_Index = list(range(1, $\mathcal{N}_U+1$)) \\
        \STATE Generate an iterator that produces tuples of length ${N_u}$ representing all possible combinations of ${N_u}$ UAVs, i.e., $\{c_1^{N_u}, c_2^{N_u}, \cdot\cdot\cdot, c_{\mathcal{M}}^{N_u}\} \in \mathcal{C}$, from the list UAV\_Index: \\
        $\mathcal{C} = $itertools.combinations(UAV\_Index, ${N_u}$)
        \FOR{$j=1$ to $\mathcal{N}_u$}
            \STATE Obtain the CQI submatrix $\mathbf{H}_{t,c_j^{N_u}} \in \mathbf{H}_t$ corresponding to the combination $c_j^{N_u}$
            \STATE Obtain the weight vector:\\
            $ \mathbf{w}_{t, c_j^{N_u}} = \frac{\mathbf{H}_{t,c_j^{N_u}}^H}{||\mathbf{H}_{t,c_j^{N_u}}^H||}$ \\
            \STATE Calculate SINR: $F$
            \IF{$F > F_{Max}$}
                \STATE $C_{opt}^{N_u} = c_j^{N_u}$
                \STATE $F_{Max} = F$
            \ENDIF
        \ENDFOR
        \RETURN $C_{opt}^{N_u}$
        \WHILE{$i \leq I_Q$}
        \STATE Apply $H_d^{\bigotimes\log_2(K_u^{N_u})}$ to the first $\log_2(K_u^{N_u})$ qubits to obtain the initial quantum superposition:\\
        $Q\left( {\left| 0 \right.\rangle ^{ \otimes \log _2 \left( {K_u^{N_u } } \right)} } \right)$
        \STATE Apply $H_dX$ to the last qubit to retain the superposition state introduced by $H_d$ \\
        \FOR{$j=1$ to $\sqrt{2^{\log_2(K_u^{N_u})}}$}
            \STATE Apply the phase inversion operation: $U_f(I\bigotimes H_d)$
            \STATE Apply the inversion about the mean operation: $-I + 2A$\\
        \ENDFOR
        \STATE Measure the qubits
        \STATE $i \leftarrow i + 1$
        \ENDWHILE
        \RETURN $\mathbf{E}_{\rm opt} $
        \end{algorithmic}
        \label{ALGO-I}
\end{algorithm}

Recall that $\mathcal{S}_{\rm opt} = \{\mathcal{S}_{\rm opt}^{(1)}, \mathcal{S}_{\rm opt}^{(2)}, \cdot\cdot\cdot, \mathcal{S}_{\rm opt}^{(\mathcal{S}_M)}\} \in \mathcal{S}; 1 \leq \mathcal{S}_M \leq K_u^{N_u}$, is the solution space that contains all the possible solutions to the search problem. Following this, and as illustrated in Algorithm \ref{ALGO-I}, in the proposed QSUB and Q-P-LL frameworks, it is important to apply the quantum oracle with the oracle qubit initially in the state $(|0\rangle - |1\rangle)/\sqrt{2}$. As illustrated in Fig. \ref{FIG-4}, if any UAS state $s$ is not the optimal solution to the search problem, then applying the oracle to the state $s$ does not change $(|0\rangle - |1\rangle)/\sqrt{2}$. On the contrary, if $s$ is a solution to the problem of finding the optimal UASs locations, then $|0\rangle$ and $|1\rangle$ are interchanged by the action of the quantum oracle, generating a final state $-|s\rangle(|0\rangle - |1\rangle)/\sqrt{2}$. The action of the quantum oracle can then be defined as
\begin{equation}
\left| s \right.\rangle \left( {\frac{{\left| {0\rangle  - \left| 1 \right.\rangle } \right.}}{{\sqrt 2 }}} \right) \to \left. {\left( { - 1} \right)^{f\left( s \right)} } \right|s\rangle \left( {\frac{{\left| {0\rangle  - \left| 1 \right.\rangle } \right.}}{{\sqrt 2 }}} \right)
\label{EQ-6}.
\end{equation}

The pictorial representation of the proposed QSUB is illustrated in Fig. \ref{FIG-6}. Two-dimensional geometric section of the Hilbert space. The vectors shown in the figure are geometric representations of the quantum states in a Hilbert space. $|\omega\rangle$ represents the present superposition state of the quantum system whereas $|s\rangle$ is the desired quantum state (solution). $U_x(s_j)$ represents the unitary operation which reflect the state $s_j$ over the axis denoted by the vector $x$. The dimension of the Hilbert space is $K_u^{N_u}$. The objective is to maximize the probability of measuring the desired state $|s\rangle$. The state $|\omega\rangle$ includes all the computational basis states with equal coefficients, such that, all the computational bases have equal probabilities of being measured.

\section{{Q-P-LL: Optimized QSUB}}
Next, we formulate an optimization problem to minimize the error in QSUB. It can be written as
{\begin{equation}
\Lambda  = \mathop {\min }\limits_{b \in \mathbb{R}^{K_o \in S} } f_{\rm QSUB}\left( b \right)
\end{equation}}
where $\mathbb{R}$ represents the set of real numbers. $K_o \in \mathbb{Z}^+$ denotes the number of optimization variables in vector $b \in \mathbb{R}^{K_o}$. $\mathbb{Z}^+$ is the set of positive integers. $f_{QSUB}: \mathbb{R}^{K_o} \rightarrow \mathbb{R}$. Similar to the Nelder-Mead optimization algorithm, the proposed optimization problem maintains a simplex $\mathcal{P}$ containing $K_o + 1$ vertices, with each vertex in $\mathcal{P}$ having associated with it the value of the objective function at the respective vertex \cite{kelley1999iterative}. 

In the optimization of QSUB, the vertices, $b_j \in \mathbb{R}^{K_o}, j = \{1, 2, \cdot\cdot\cdot, K_o + 1\}$, are sorted according to their respective objective function values
{\begin{equation}
f_{QSUB} \left( {b_1 } \right) \le f_{QSUB} \left( {b_2 } \right) \le  \cdot  \cdot  \cdot  \le f_{QSUB} \left( {b_{K_o  + 1} } \right)
\end{equation}}

At each iteration value, the optimization problem attempts to replace the worst point $b_{K_o + 1}$ by first computing the centroid $B_C$, expressed as
{\begin{equation}
B_C  = \frac{1}{{K_o }}\sum\limits_{j = 1}^{K_o } {b_j }
\end{equation}}
whereas, the trial vertices $b\left(\eta\right) \in \mathbb{R}^{K_o}$ can then readily be calculated as
{\begin{equation}
b\left( \eta  \right) = \left( {1 + \eta } \right)B_C  - \eta b_{K_o  + 1}
\label{EQ-17A}
\end{equation}}
where $\eta \in \mathbb{R}$ denotes the coefficient associated with a particular step in the optimization process. It is important to note that, as represented in the Algorithm \ref{ALGO-II}, (\ref{EQ-17A}) corresponds to the reflection, expansion, contraction, and shrink steps with respective coefficient $r_c \in \mathbb{R}, e_c \in \mathbb{R}, c_c \in \mathbb{R}, s_c \in \mathbb{R}$ \cite{wyers2013bounded}.

As depicted in Algorithm \ref{ALGO-II}, if the reflection, expansion, and contraction steps fail to obtain an improvement on the worst point $b_{K_o + 1}$, then Algorithm \ref{ALGO-II} performs a shrink step, with shrink vertices obtained as
{\begin{equation}
\mathcal{P}\left( k \right) \leftarrow \mathcal{P}\left( 0 \right) + s_c \left( {\mathcal{P}\left( k \right) - \mathcal{P}\left( 0 \right)} \right)
\end{equation}}
However, if the reflection, expansion, and contraction steps are successful in replacing the worst point $b_{K_o + 1}$, then the average simplex objective function value can readily be obtained as
{\begin{equation}
\bar f_{Q - P - LL}  = \frac{1}{{K_o  + 1}}\sum\limits_{j = 1}^{K_o  + 1} {f_{QSUB} \left( {b_j } \right)}
\end{equation}}

\begin{algorithm}
	\renewcommand{\algorithmicrequire}{\textbf{Input:}}
	\renewcommand{\algorithmicensure}{\textbf{Output:}}
	\caption{{Q-P-LL: Optimized QSUB}}
        \textbf{Input}: Objective function: $f\left( {\mathbf{E},\mathbf{\Theta}} \right)$; Oracle $\mathcal{O}$ to perform the transformation: $O|s\rangle q\rangle  = |s\rangle q \oplus f\left( s \right)\rangle$ \\
        \textbf{Input}: Reflection, Expansion, Contraction, and Shrink coefficients: $r_c, e_c, c_c, s_c$; Population of simplex $\mathcal{P}$\\
        \textbf{Input}: Tolerance: $\mathcal{T}$;  Iterations: $I_{Q}$; Simplex points: $N_s$ \\
        \textbf{Initialize}: Quantum states:  $\left| {\mathbf{0}\rangle  = } \right.\left| 0 \right.\rangle ^{ \otimes \log_2({K_u^{N_u }}) } \left| {0\rangle } \right.$\\
        \begin{algorithmic}[1]
        \STATE Apply $H_d^{\bigotimes\log_2(K_u^{N_u})}$ to the first $\log_2(K_u^{N_u})$ qubits to obtain the initial quantum superposition:
        $Q\left( {\left| 0 \right.\rangle ^{ \otimes \log _2 \left( {K_u^{N_u } } \right)} } \right)$
        \STATE Apply $H_dX$ to the last qubit to retain the superposition \\
        \FOR{$j=1$ to $\sqrt{2^{\log_2(K_u^{N_u})}}$}
            \STATE Apply the phase inversion operation: $U_f(I\bigotimes H_d)$
            \STATE Apply the inversion about the mean operation: $-I + 2A$\\
        \ENDFOR
        \WHILE{$i \leq I_{Q}$}
        \STATE $\mathcal{P} \leftarrow {\rm sort}\left(\mathcal{P}\right)$: such that\\
        $f\left( {\mathbf{E}_0^{(i)},\mathbf{\Theta}} \right) \leq \cdot\cdot\cdot \leq
        f\left( {\mathbf{E}_j^{(i)},\mathbf{\Theta}} \right) \leq \cdot\cdot\cdot \leq
        f\left( {\mathbf{E}_{N_s}^{(i)},\mathbf{\Theta}} \right)$
        \STATE \textbf{Evaluate:} $m_\mathcal{P}  \leftarrow \frac{1}{{N_s }}\sum\limits_{j = 0}^{N_s  - 1} {\mathcal{P}\left( j \right)}$
        \STATE \textbf{Evaluate:} $r_\mathcal{P}  \leftarrow m_\mathcal{P}  + r_c \left( {m_\mathcal{P}  - \mathcal{P}\left( {N_s - 1} \right)} \right)$
        \STATE \IF{$f\left( {\mathbf{E}_0^{(i)},\mathbf{\Theta}} \right) < f\left( {\mathbf{E}_{r_\mathcal{P} }^{(i)},\mathbf{\Theta}} \right) < f\left( {\mathbf{E}_{N_s - 1 }^{(i)},\mathbf{\Theta}} \right)$}
        \STATE $\mathcal{P}\left( {N_s - 1} \right) \leftarrow r_\mathcal{P}$
        \ELSE
        \IF{$f\left( {\mathbf{E}_{r_\mathcal{P} }^{(i)},\mathbf{\Theta}} \right) \le f\left( {\mathbf{E}_0^{(i)},\mathbf{\Theta}} \right)$}
        \STATE \textbf{Evaluate:} $e_\mathcal{P} \leftarrow r_\mathcal{P} + c_c(r_\mathcal{P} - m\mathcal{P})$
        \IF{{$f\left( {\mathbf{E}_{e_{\mathcal{P}}}^{(i)},\mathbf{\Theta}} \right) \le f\left( {\mathbf{E}_{r_{\mathcal{P}}}^{(i)},\mathbf{\Theta}} \right)$}}
        \STATE $\mathcal{P}\left( {N_s - 1} \right) \leftarrow e_\mathcal{P}$ 
        \ELSE
        \STATE $\mathcal{P}\left( {N_s - 1} \right) \leftarrow r_\mathcal{P}$
        \ENDIF
        \ELSE
        \STATE $t_f \leftarrow {\rm True}$
        \IF{$f\left( {\mathbf{E}_{r_{\mathcal{P}}}^{(i)},\mathbf{\Theta}} \right) \geq f\left( {\mathbf{E}_{N_s - 2}^{(i)},\mathbf{\Theta}} \right)$}
        \STATE \textbf{Evaluate:} $c_\mathcal{P} \leftarrow e_c r_\mathcal{P} + \left(1 - e_c\right)m_\mathcal{P}$
        \IF{$f\left( {\mathbf{E}_{c_{\mathcal{P}}}^{(i)},\mathbf{\Theta}} \right) < f\left( {\mathbf{E}_{r_\mathcal{P} }^{(i)},\mathbf{\Theta}} \right)$}
        \STATE $\mathcal{P}\left(N_s - 1 \right) \leftarrow c_\mathcal{P}$; $t_f \leftarrow {\rm False}$
        \ENDIF
        \ENDIF
        \IF{$t_f = {\rm True}$}
        \FOR{$k = N_s - 1$ down to $1$}
        \STATE $\mathcal{P}(k) \leftarrow \mathcal{P}(0) + s_c \left(\mathcal{P}(k) - \mathcal{P}(0) \right)$
        \ENDFOR
        \ENDIF
        \ENDIF
        \ENDIF
        \IF{$\left(\left|f\left( {\mathbf{E}^{(i)},\mathbf{\Theta}} \right) - f\left( {\mathbf{E}^{(i-1)},\mathbf{\Theta}} \right)\right| \leq \mathcal{T}\right)$}
        \RETURN $\mathbf{E}_{\rm opt} = \mathbf{E}^{(i)}$
        \ENDIF
        \ENDWHILE
        \RETURN $\mathbf{E}_{\rm opt}$
        \end{algorithmic}
        \label{ALGO-II}
\end{algorithm}

\begin{figure}
    \centering
    \begin{subfigure}[b]{0.50\textwidth}
        \centering
        \includegraphics[height=1.4in]{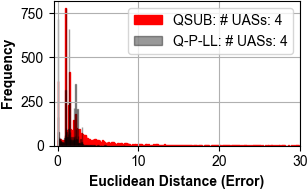}
        \caption{Performance comparison $(\sigma_{AoA} = 0)$.}
        \label{FIG-7A}
    \end{subfigure}%
    
    \begin{subfigure}[b]{0.50\textwidth}
        \centering
        \includegraphics[height=1.4in]{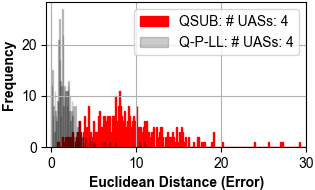}
        \caption{Performance comparison $(\sigma_{AoA} = 5)$.}
        \label{FIG-7B}
    \end{subfigure}
    
    \caption{Q-P-LL versus QSUB: Distribution of the Euclidean error.}
    \label{FIG-7}
\end{figure}

\begin{figure}
    \centering
        \includegraphics[height=2.0in,width=2.7in]{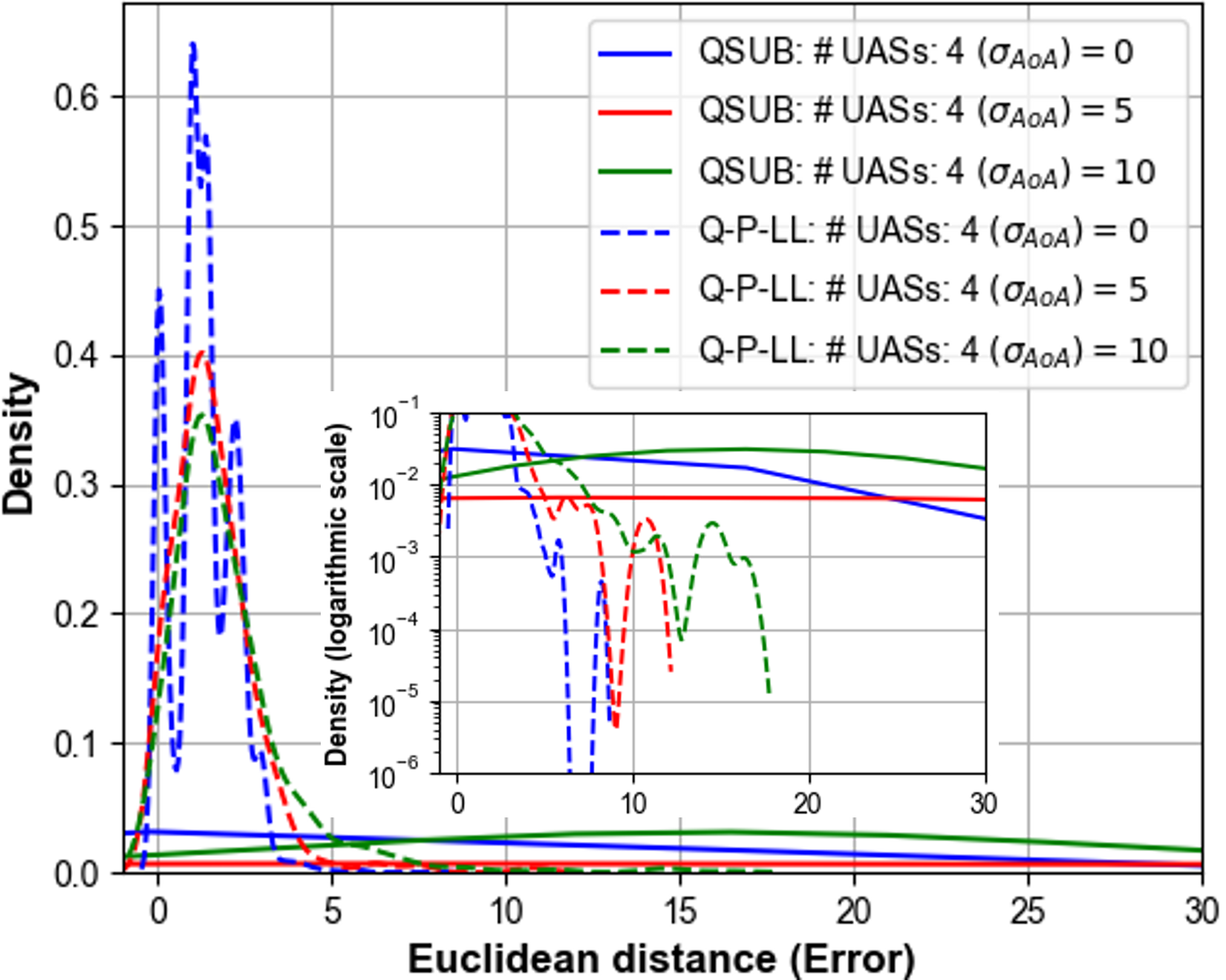}
    \caption{Density.}
    \label{FIG-8}
\end{figure}

\begin{figure}
    \centering
        \includegraphics[width=2.8in, height=2.0in]{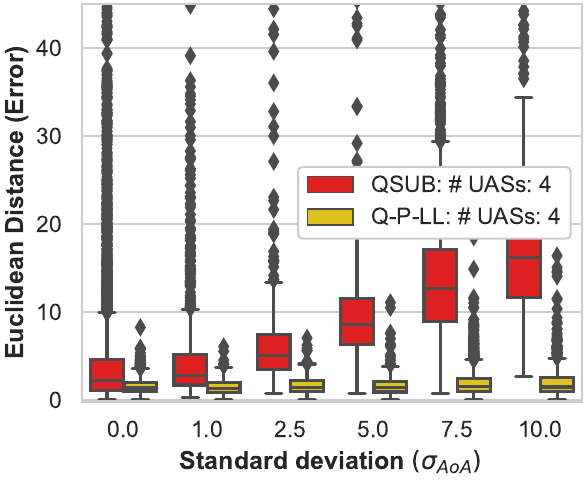}
    \caption{Performance comparison of Q-P-LL relative to QSUB.}
    \label{FIG-9}
\end{figure}

Algorithm \ref{ALGO-II} illustrates the Q-P-LL framework. 
\begin{figure}
    \centering
        \includegraphics[width=2.8in, height=2.0in]{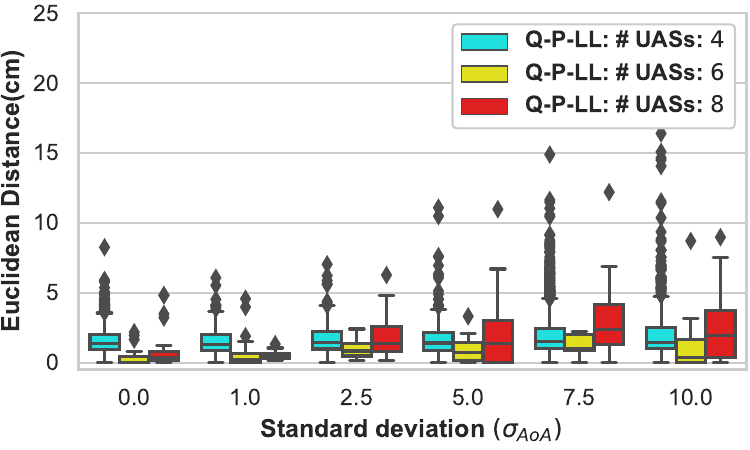}
    \caption{Performance illustration of the proposed Q-P-LL for different numbers of UASs.}
    \label{FIG-10}
\end{figure}

\section{Complexity Analysis of the Q-P-LL}\label{SECTION-III}
To find the optimal state $|\varphi_{\rm opt}\rangle$, we apply the quantum oracle $\mathcal{O}$ for the transformation such that $\mathcal{O}$ generates a phase shift of $-1$ to $|\varphi_{\rm opt}\rangle$ whereas leaves all other states $\{ s | s \in \mathcal{S}, s \neq |\varphi_{\rm opt}\rangle\}$ invariant. With oracle $\mathcal{O}$ applies exactly $b$ times with exactly $b$ number of unitary operations $U_1, U_2, \cdot\cdot\cdot, U_b$ interleaved between the oracle operations.

Let $\delta_b$ be the deviation caused by the oracle after $b$ steps, and is defined as
\begin{equation}
\delta _b  \equiv \sum\limits_j {\left\| {\left| \varphi  \right._b^j \rangle  - \left| {\varphi _b \rangle } \right.} \right\|^2 }
\end{equation}
where $\left| \varphi  \right._b^j \rangle$ is given by
\begin{equation}
\left| \varphi  \right._b^j \rangle  = U_b \mathcal{O}U_{b - 1} \mathcal{O} \cdot  \cdot  \cdot U_1 \mathcal{O}\left| \varphi  \right._{{\mathop{\rm int}} } \rangle
\end{equation}
and
\begin{equation}
\left| \varphi  \right._b \rangle  = U_b U_{b - 1}  \cdot  \cdot  \cdot U_1 \left| \varphi  \right._{{\mathop{\rm int}} } \rangle
\end{equation}
Notifying that,
\begin{equation}
    \begin{split}
& \delta _{b + 1}  \equiv \sum\limits_j {\left\| { {\mathcal{O}_j \varphi _b^j \rangle }  - \left| {\varphi _b \rangle } \right.} \right\|^2 }  \\ 
& \hspace{0.7cm} = \sum\limits_j {\left\| { {\mathcal{O}_j \left( {\left| \varphi  \right._b^j \rangle  - \left| {\varphi _b \rangle } \right.} \right)} + \left( {\mathcal{O}_j  - I} \right)\left| {\varphi _b \rangle } \right.} \right\|^2 }  \\ 
    \end{split}
\end{equation}
Utilizing the inequality $\left\| {x + y} \right\|^2  \le \left\| x \right\|^2  + \left\| y \right\|^2  + 2\left\| x \right\|\left\| y \right\|$, it can readily be shown that $\rho_{b+1}$ satisfies (\ref{EQ-21}).
\begin{figure*}
    \begin{equation}
\delta _{b + 1}  \le \sum\limits_j {\left( {\left\| {\left| \varphi  \right._b^j \rangle  - \left| {\varphi _b \rangle } \right.} \right\|^2  + 4\left\| {\left| \varphi  \right._b^j \rangle  - \left| {\varphi _b \rangle } \right.} \right\|\left| {\langle s_j \left| {\varphi _b \rangle } \right.} \right| + 4\left| {\langle \varphi _b \left| {s_j \rangle } \right.} \right|^2 } \right)}
\label{EQ-21}
\end{equation}
\end{figure*}
Utilizing the Cauchy-Schwarz inequality, the inequality can be expressed as illustrated in (\ref{EQ-22}).
\begin{figure*}
    \begin{equation}
\delta _{b + 1}  \le \sum\limits_j {\left\| {\left| {\mathcal{O}_j \varphi _b^j \rangle } \right. - \left| {\varphi _b \rangle } \right.} \right\|^2 }  + 4\sqrt {\sum\limits_j {\left\| {\left| {\mathcal{O}_j \varphi _b^j \rangle } \right. - \left| {\varphi _b \rangle } \right.} \right\|^2 } } \sqrt {\sum\limits_k {\left| {\langle \varphi _b \left| {s_k \rangle } \right.} \right|^2 } }  + 4
    \label{EQ-22}.
    \end{equation}
\end{figure*}
Substituting $\left. {\langle s} \right|$ by $\exp(i\theta \left| s \right.\rangle)$ yields $\left. {\langle s_j } \right|\varphi _b^j \rangle  = \left| {\left. {\langle s_j } \right|\varphi _b^j \rangle } \right|$. Following this, it can shown that
\begin{equation}
\left\| {\varphi _b^j \rangle  - s_j } \right\|^2  = 2 - 2\left| {\langle s_j \left| {\varphi _b^j } \right.\rangle } \right| \le 2 - \sqrt 2
\label{EQ-23}.
\end{equation}
It yields $\sum\limits_j {\left\| { {\left| \mathcal{O}_j \right. \varphi _b^j \rangle }  - \left| {s_j \rangle } \right.} \right\|^2 }  \le \sqrt 2 \left( {\sqrt 2  - 1}. \right)K_u^{N_u } $. Following this, we $\delta_b$ can be expressed as (\ref{EQ-24}).

    \begin{equation}
    \begin{split}
& \delta _b  = \sum\limits_j {\left\| {\left( {\left| \varphi  \right._b^j \rangle  - \left| {s_j \rangle } \right.} \right) + \left( {\left| {s_j \rangle } \right. - \left| {\varphi _b \rangle } \right.} \right)} \right\|^2 }  \\ 
& \hspace{0.3cm} \ge \sum\limits_j {\left\| {\left| \varphi  \right._b^j \rangle  - \left| {s_j \rangle } \right.} \right\|^2 }  + \sum\limits_j {\left\| {\left| {s_j \rangle } \right. - \left| {\varphi _b \rangle } \right.} \right\|^2 }  \\
& \hspace{0.3cm}- 2\sum\limits_j {\left\| {\left| \varphi  \right._b^j \rangle  - \left| {s_j \rangle } \right.} \right\|\left\| {\left| {s_j \rangle } \right. - \left| {\varphi _b \rangle } \right.} \right\|}  \\
    \end{split}
    \label{EQ-24}.
\end{equation}
Application of the Cauchy-Schwarz inequality yields
\begin{equation}
    \begin{split}
& \sum\limits_j {\left\| {\left| \varphi  \right._b^j \rangle  - \left| {s_j } \right.\rangle } \right\|\left\| {\left| {s_j \rangle } \right. - \left| {\varphi _b \rangle } \right.} \right\|}  \\ 
& \hspace{0.1cm}  \le \sqrt {\sum\limits_j {\left\| {\left| \varphi  \right._b^j \rangle  - \left| {s_j \rangle } \right.} \right\|^2 \sum\limits_j {\left\| {\left| {s_j \rangle } \right. - \left| {\varphi _b \rangle } \right.} \right\|^2 } } }
    \end{split}
    \label{EQ-25}
\end{equation}
Following (\ref{EQ-25}), it yields
\begin{equation}
\delta _b  \ge \left( {\sum\limits_j \left\|{ {\left| {s_j \rangle } \right. - \left| {\varphi _b \rangle } \right.} }\right\|^2  - \sum\limits_j {\left\| {\left| \varphi  \right._b^j \rangle  - \left| {s_j \rangle } \right.} \right\|^2 } } \right)^2
\label{EQ-26}.
\end{equation}
For any normalized quantum state vector $|\varphi_b \rangle$ and set of $K_u^{N_u}$ orthonormal basis vectors $|s\rangle$, the application of Cauchy-Schwarz inequality yields
\begin{equation}
\sum\limits_j {\left\| {\left| \varphi  \right._b \rangle  - \left| {s_j \rangle } \right.} \right\|^2 }  \ge 2\sqrt {K_u^{N_u } } \left( {\sqrt {K_u^{N_u } }  - 1} \right)
\label{EQ-27}.
\end{equation}
Combining (\ref{EQ-23}) and (\ref{EQ-27}), yields $\delta_b \geq cK_u^{N_u}$ for large $K_u^{N_u}$. Here $c$ is a constant which yields $c < \left(\sqrt{2} - \sqrt{(2 - \sqrt{2})}\right)^2$. It follows
\begin{equation}
b \ge \frac{1}{2}\sqrt {K_u^{N_u } \left( {\sqrt 2  - \sqrt {2 - \sqrt 2 } } \right)^2 }
\label{EQ-28}.
\end{equation}
For an observation to generate a solution to the position-locked loop problem with a probability of at least one-half, we must have $
\left| {\langle s_j \left| {\left| \varphi  \right._b^j } \right.\rangle } \right|^2  \ge \frac{1}{2}; \forall j$. Therefore, to achieve a success probability of the Q-P-LL at one-half for locking the correct position to the problem of beam re-forming, the oracle must be called $\Delta(K_u^{N_u})$ times.

\section{Performance Analysis}
The tolerance $\mathcal{T}$ is set to $10^{-3}$. The operating frequency is set to $3.5$ GHz. In obtaining the results, while selecting the optimal UASs, the spacing between the selected UAVs can be set a bit larger to gain higher beamforming resolution but is ensured that there should not be any aliasing.  
Figure \ref{FIG-7} illustrates the performance comparison of the two proposed frameworks, i.e., QSUB and Q-P-LL for different $\sigma_{AoA}$ values. The distribution of the Euclidean distance $r_e$ is analyzed for $N_u = 4$. As illustrated, the limitation of QSUB is that it can only handle cases with negligible estimation error with $\sigma_{AoA}   \rightarrow  0$. Compared to the QSUB, the Q-P-LL is especially useful for the beam re-forming under the influence of the AoA estimation error.

In Fig. \ref{FIG-8}, the probability of $r_e$ is plotted for QSUB and Q-P-LL for different $\sigma_{AoA}$ values. The curves are also shown on a logarithmic scale from which it is evident that the Q-P-LL has a definite upper bound on the $r_e$ corresponding to $\sigma_{AoA}$ values.

To analyze the proposed QSUB and Q-P-LL further and make it easy to identify patterns and outliers, Fig. \ref{FIG-9} shows the spread (variability) and central tendency of $r_e$ as a function of $\sigma_{AoA}$. As can be seen clearly, as the value of $\sigma_{AoA}$ increases, the spread of $r_e$ increases significantly. Importantly, it is also evident that Q-P-LL performs well, and when compared to QSUB, does not vary much with $\sigma_{AoA}$.

\begin{figure}
    \centering
    \begin{subfigure}[b]{0.51\textwidth}
        \centering
        \includegraphics[height=1.7in]{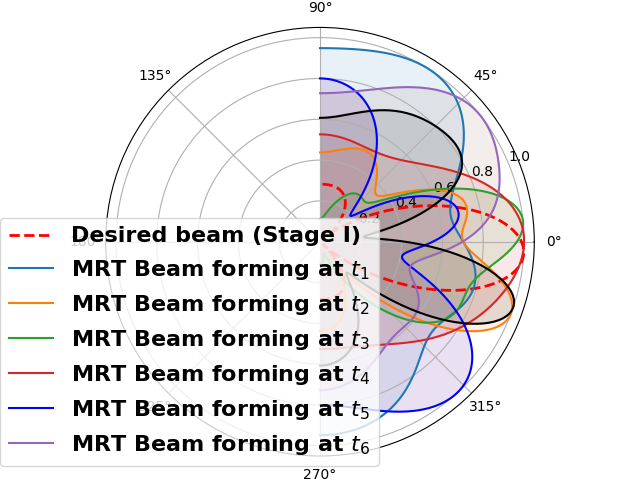}
        \caption{Beamforming traces after Stage I at different instants using conventional MRT scheme (without QSUB and Q-P-LL applied) $(\sigma_{AoA} = 0)$.}
        \label{FIG-11A}
    \end{subfigure}%
    
    \begin{subfigure}[b]{0.51\textwidth}
        \centering
        \includegraphics[height=1.7in]{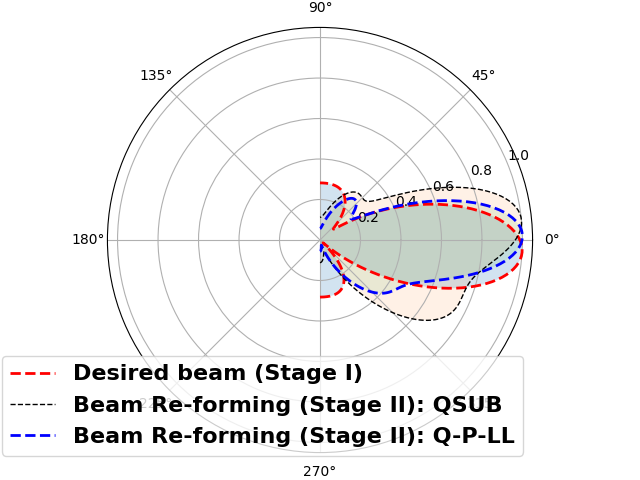}
        \caption{Performance comparison $(\sigma_{AoA} = 0)$.}
        \label{FIG-11B}
    \end{subfigure}

    \begin{subfigure}[b]{0.51\textwidth}
        \centering
        \includegraphics[height=1.6in]{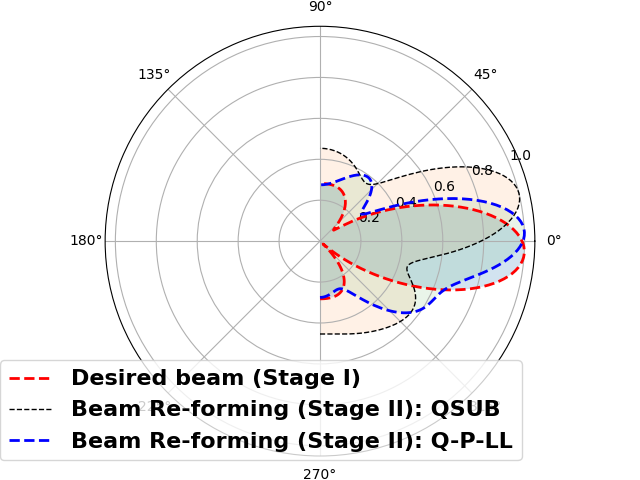}
        \caption{Performance comparison $(\sigma_{AoA} = 2.5)$.}
        \label{FIG-11C}
    \end{subfigure}

    \begin{subfigure}[b]{0.51\textwidth}
        \centering
        \includegraphics[height=1.7in]{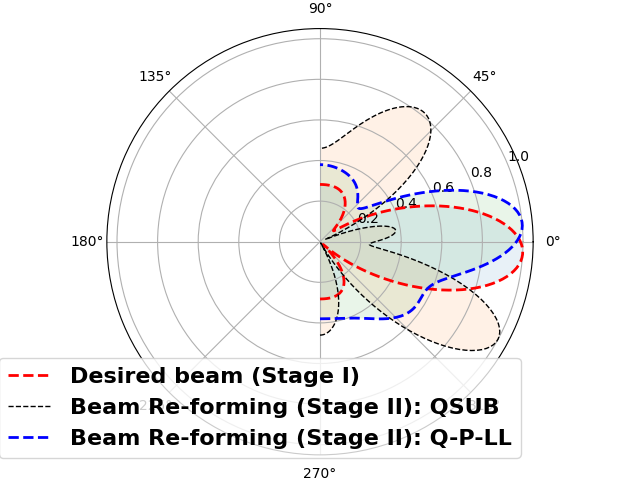}
        \caption{Performance comparison $(\sigma_{AoA} = 5)$.}
        \label{FIG-11D}
    \end{subfigure}
  
    \caption{Beam re-forming and performance comparison of QSUB and Q-P-LL for different $\sigma_{AoA}$.}
    \label{FIG-11}
\end{figure}

The performance of the Q-P-LL as a function of the number of UASs $N_u$ is depicted in Fig. \ref{FIG-10}. Interestingly, it can be observed that, for small values of $\sigma_{AoA}$, the Euclidean distance decreases with increasing the number of UASs, however, as the standard deviation increases beyond $2.5$, the error increases significantly with the number of UASs. Another observation that can be inferred from the results is that although the error increases with the number of USAs, the number of outliers also decreases.

The Q-P-LL is tested in terms of accuracy in the produced radiation pattern. The results are generated on $5682$ trials. The statistical analysis of the results derived from this test is presented in Table \ref{TABLE-II}. From the results obtained, it can be observed that the Q-P-LL implementation shows impressive performance regarding the main lobe steering as the mean value of the main lobe divergence is significantly less even for high $\sigma_{AoA}$ values. Interestingly, it can be seen that the performance improves as the number of UAS increases from $4$ to $6$, however, it decreases, as the number of UAS increases from $6$ to $8$. This can also be seen in terms of Euclidean distance error as shown in Fig. \ref{FIG-10}. This phenomenon can be attributed to the fundamental problem of quantum computation of deciding how many copies of an unknown mixed quantum state are necessary and sufficient to determine the state. In quantum computation, this is due to the problem of finding an optimal quantum measurement for the generalized quantum state discrimination task, which includes the problem of finding an optimal measurement for maximum success probability \cite{nakahira2016finding}. As can be seen, our proposed Q-P-LL provides a promising and acceptable solution to the beamforming impairments in IoT communication networks.

\begin{table}
	\renewcommand{\arraystretch}{1.3}
	\caption{Q-P-LL Performance measurement: Beam divergence of the main lobe and side lobe.}
	\label{TABLE-II}
	\centering
	\begin{tabular}{ |l|l|l| p{4cm} } 
		\hline
		\textbf{$\#$UAS} & \textbf{Divergence of main lobe} & \textbf{Divergence of nulls}\\
		\hline
		$ 4 (\hat{\sigma}_{AoA} = 0) $ &  $ 1.38^\circ $ &  $ 1.90^\circ $\\
        \hline
		$ 4 (\hat{\sigma}_{AoA} = 1)$ &  $ 2.98^\circ $ &  $ 2.12^\circ $\\
        \hline
		$ 4 (\hat{\sigma}_{AoA} = 2.5)$ &  $ 4.01^\circ $ &  $ 2.97^\circ $\\
        \hline
        \hline
		$ 6 (\hat{\sigma}_{AoA} = 0)$ & $ 0.87^\circ $ &  $ 0.98^\circ $ \\
		\hline
        $ 6 (\hat{\sigma}_{AoA} = 1)$ & $ 1.03^\circ $ &  $ 1.05^\circ $ \\
		\hline
        $ 6 (\hat{\sigma}_{AoA} = 2.5)$ & $ 1.81^\circ $ &  $ 2.76^\circ $ \\
		\hline
        \hline
        $ 8 (\hat{\sigma}_{AoA} = 0)$    &    $ 1.03^\circ $ &  $ 2.01^\circ $ \\
        \hline
        $ 8 (\hat{\sigma}_{AoA} = 1)$    &    $ 1.07^\circ $ &  $ 3.72^\circ $ \\
        \hline
        $ 8 (\hat{\sigma}_{AoA} = 2.5)$    &    $ 5.23^\circ $ &  $ 7.63^\circ $ \\
        \hline
        \hline
    \end{tabular}
\end{table}

The beam re-forming comparison of QSUB and Q-P-LL is illustrated in Fig. \ref{FIG-11} for different $\sigma_{AoA}$. Note that, the $AoA$ estimation error $\sigma_{AoA}$ can be reflected by beam distortion. However, comparing QSUB with Q-P-LL, we observe that the received beam for Q-P-LL is more concentrated around the receiver, thereby, validating the superior performance of Q-P-LL. A comparison with the conventional maximum ratio transmission (MRT) scheme is illustrated in Fig. \ref{FIG-11A}. As depicted in Fig. \ref{FIG-11A}, the conventional MRT alone is not suitable for the UAS network as the hovering distorts the beam.

\begin{figure}
    \centering
        \includegraphics[width=0.93\linewidth]{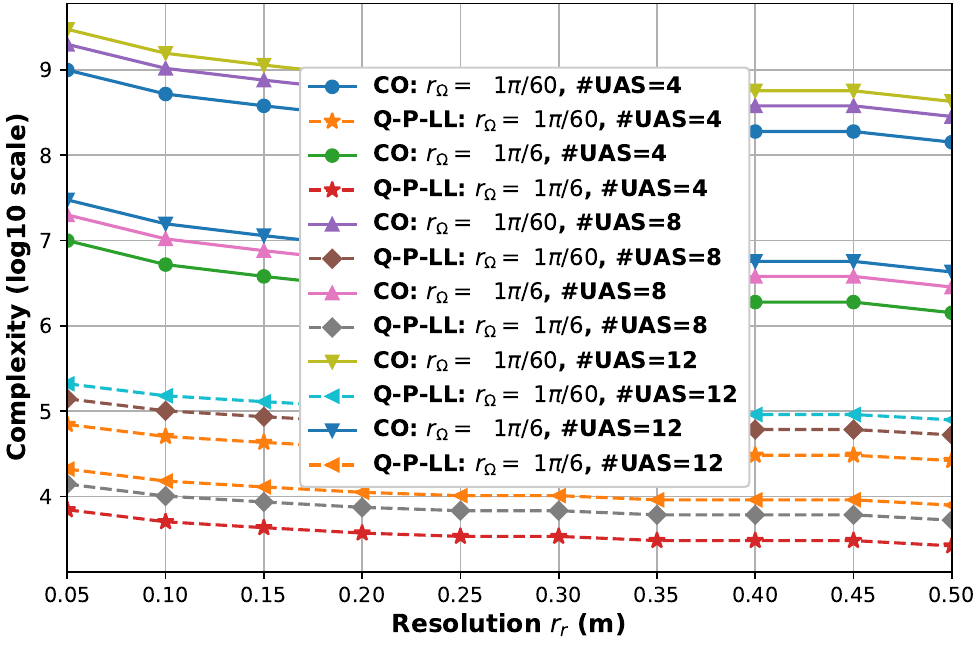}
    \caption{Complexity analysis of Q-P-LL compared to the classical oracle for different $N_u$ and $r_\Omega; \Omega \in \{\theta, \phi\}$.}
    \label{FIG-12}
\end{figure}

The complexity analysis of the Q-P-LL is presented in Fig. \ref{FIG-12}. The Q-P-LL is characterized by a lower complexity when compared with the classical oracle. As illustrated, Q-P-LL significantly reduces the computational complexity, and this effect is more pronounced as the number of UAS increases. 

\section{Conclusion}
This work considered a quantum computation approach to the problem of collaborative beamforming and beam re-forming for a distributed UAS-assisted network. For the provision of opportunistic access to achieve higher SINR performance, we first proposed a QSUB framework that relies on Grover’s quantum search algorithm that identifies the optimal links to generate beamforming and then adapts the UASs according to their random states due to the hovering. Although QSUB was shown to perform well under random hovering conditions, it was found that the performance is sensitive to the AoA estimation error. Finally, to overcome the impact of the AoA estimation error, we presented a new optimized framework Q-P-LL, and it was demonstrated that Q-P-LL significantly minimizes the impact of the AoA estimation error without increasing the complexity. Both frameworks, QSUB and Q-P-LL, overcome the adverse impact of the random hovering of the UASs and do not require channel information.



\ifCLASSOPTIONcaptionsoff
  \newpage
\fi

\bibliographystyle{IEEEtran}
\bibliography{IEEEabrv,Bibliography,references_ying}

\begin{thebibliography}{10}
\providecommand{\url}[1]{#1}
\csname url@rmstyle\endcsname
\providecommand{\newblock}{\relax}
\providecommand{\bibinfo}[2]{#2}
\providecommand\BIBentrySTDinterwordspacing{\spaceskip=0pt\relax}
\providecommand\BIBentryALTinterwordstretchfactor{4}
\providecommand\BIBentryALTinterwordspacing{\spaceskip=\fontdimen2\font plus
\BIBentryALTinterwordstretchfactor\fontdimen3\font minus \fontdimen4\font\relax}
\providecommand\BIBforeignlanguage[2]{{%
\expandafter\ifx\csname l@#1\endcsname\relax
\typeout{** WARNING: IEEEtran.bst: No hyphenation pattern has been}%
\typeout{** loaded for the language `#1'. Using the pattern for}%
\typeout{** the default language instead.}%
\else
\language=\csname l@#1\endcsname
\fi
#2}}

\bibitem{chettri2019comprehensive}
L.~Chettri and R.~Bera, ``A comprehensive survey on internet of things (iot) toward 5g wireless systems,'' \emph{IEEE Internet of Things Journal}, vol.~7, no.~1, pp. 16--32, 2019.

\bibitem{lopez2021massive}
O.~L. L{\'o}pez, H.~Alves, R.~D. Souza, S.~Montejo-S{\'a}nchez, E.~M.~G. Fern{\'a}ndez, and M.~Latva-Aho, ``Massive wireless energy transfer: Enabling sustainable iot toward 6g era,'' \emph{IEEE Internet of Things Journal}, vol.~8, no.~11, pp. 8816--8835, 2021.

\bibitem{bhatia2020quantum}
M.~Bhatia and S.~K. Sood, ``Quantum computing-inspired network optimization for iot applications,'' \emph{IEEE Internet of Things Journal}, vol.~7, no.~6, pp. 5590--5598, 2020.

\bibitem{caleffi2023beyond}
M.~Caleffi, K.~Simonov, and A.~S. Cacciapuoti, ``Beyond shannon limits: Quantum communications through quantum paths,'' \emph{IEEE Journal on Selected Areas in Communications}, 2023.

\bibitem{garcia2014quantum}
J.~C. Garcia-Escartin and P.~Chamorro-Posada, ``Quantum spread spectrum multiple access,'' \emph{IEEE Journal of Selected Topics in Quantum Electronics}, vol.~21, no.~3, pp. 30--36, 2014.

\bibitem{mohanty2023quantum}
T.~Mohanty, V.~Srivastava, S.~K. Debnath, A.~K. Das, and B.~Sikdar, ``Quantum secure threshold private set intersection protocol for iot-enabled privacy preserving ride-sharing application,'' \emph{IEEE Internet of Things Journal}, 2023.

\bibitem{wang2022quantum}
C.~Wang and A.~Rahman, ``Quantum-enabled 6g wireless networks: Opportunities and challenges,'' \emph{IEEE Wireless Communications}, vol.~29, no.~1, pp. 58--69, 2022.

\bibitem{chen2020distributed}
R.~Chen, B.~Yang, and W.~Zhang, ``Distributed and collaborative localization for swarming uavs,'' \emph{IEEE Internet of Things Journal}, vol.~8, no.~6, pp. 5062--5074, 2020.

\bibitem{xiao2016enabling}
Z.~Xiao, P.~Xia, and X.-G. Xia, ``Enabling uav cellular with millimeter-wave communication: Potentials and approaches,'' \emph{IEEE Communications Magazine}, vol.~54, no.~5, pp. 66--73, 2016.

\bibitem{zhao2018beam}
J.~Zhao, F.~Gao, Q.~Wu, S.~Jin, Y.~Wu, and W.~Jia, ``Beam tracking for uav mounted satcom on-the-move with massive antenna array,'' \emph{IEEE Journal on Selected Areas in Communications}, vol.~36, no.~2, pp. 363--375, 2018.

\bibitem{liu2019multi}
L.~Liu, S.~Zhang, and R.~Zhang, ``Multi-beam uav communication in cellular uplink: Cooperative interference cancellation and sum-rate maximization,'' \emph{IEEE Transactions on Wireless Communications}, vol.~18, no.~10, pp. 4679--4691, 2019.

\bibitem{sun2022collaborative}
G.~Sun, J.~Li, A.~Wang, Q.~Wu, Z.~Sun, Y.~Liu, and S.~Liang, ``Collaborative beamforming for uav networks exploiting swarm intelligence,'' \emph{IEEE Wireless Communications}, vol.~29, no.~4, pp. 10--17, 2022.

\bibitem{arya2023distributed}
S.~Arya, Y.~Peng, J.~Yang, and Y.~Wang, ``Distributed 3d-beam reforming for hovering-tolerant uavs communication over coexistence: A deep-q learning for intelligent space-air-ground integrated networks,'' \emph{arXiv preprint arXiv:2307.09325}, 2023.

\bibitem{mozaffari2016efficient}
M.~Mozaffari, W.~Saad, M.~Bennis, and M.~Debbah, ``Efficient deployment of multiple unmanned aerial vehicles for optimal wireless coverage,'' \emph{IEEE Communications Letters}, vol.~20, no.~8, pp. 1647--1650, 2016.

\bibitem{8907440}
Z.~Xiao, H.~Dong, L.~Bai, D.~O. Wu, and X.-G. Xia, ``Unmanned aerial vehicle base station (uav-bs) deployment with millimeter-wave beamforming,'' \emph{IEEE Internet of Things Journal}, vol.~7, no.~2, pp. 1336--1349, 2020.

\bibitem{sun2020improving}
G.~Sun, X.~Zhao, G.~Shen, Y.~Liu, A.~Wang, S.~Jayaprakasam, Y.~Zhang, and V.~C. Leung, ``Improving performance of distributed collaborative beamforming in mobile wireless sensor networks: a multiobjective optimization method,'' \emph{IEEE Internet of Things Journal}, vol.~7, no.~8, pp. 6787--6801, 2020.

\bibitem{hou2023joint}
Z.~Hou, Y.~Huang, J.~Chen, G.~Li, X.~Guan, Y.~Xu, R.~Chen, and Y.~Xu, ``Joint irs selection and passive beamforming in multiple irs-uav enhanced anti-jamming d2d communication networks,'' \emph{IEEE Internet of Things Journal}, 2023.

\bibitem{shi2021novel}
X.~Shi, R.~Liu, and J.~S. Thompson, ``Novel distributed beamforming algorithms for heterogeneous space terrestrial integrated network,'' \emph{IEEE Internet of Things Journal}, vol.~9, no.~13, pp. 11\,351--11\,364, 2021.

\bibitem{rezai2021quantum}
M.~Rezai and J.~A. Salehi, ``Quantum cdma communication systems,'' \emph{IEEE Transactions on Information Theory}, vol.~67, no.~8, pp. 5526--5547, 2021.

\bibitem{chandra2021direct}
D.~Chandra, A.~S. Cacciapuoti, M.~Caleffi, and L.~Hanzo, ``Direct quantum communications in the presence of realistic noisy entanglement,'' \emph{IEEE Transactions on Communications}, vol.~70, no.~1, pp. 469--484, 2021.

\bibitem{9547829}
B.~Ning, Z.~Chen, Z.~Tian, C.~Han, and S.~Li, ``A unified 3d beam training and tracking procedure for terahertz communication,'' \emph{IEEE Transactions on Wireless Communications}, vol.~21, no.~4, pp. 2445--2461, 2022.

\bibitem{10161729}
W.~O.~F. Carvalho, E.~Moncada-Villa, and J.~R. Mejía-Salazar, ``Wireless at the nanoscale: Toward magnetically tunable beam steering,'' \emph{IEEE Transactions on Antennas and Propagation}, vol.~71, no.~9, pp. 7473--7479, 2023.

\bibitem{lai2022learning}
C.-Y. Lai and H.-C. Cheng, ``Learning quantum circuits of some t gates,'' \emph{IEEE Transactions on Information Theory}, vol.~68, no.~6, pp. 3951--3964, 2022.

\bibitem{nielsen2010quantum}
M.~A. Nielsen and I.~L. Chuang, \emph{Quantum computation and quantum information}.\hskip 1em plus 0.5em minus 0.4em\relax Cambridge university press, 2010.

\bibitem{jordan2018quantum}
S.~P. Jordan and Y.-K. Liu, ``Quantum cryptanalysis: shor, grover, and beyond,'' \emph{IEEE Security \& Privacy}, vol.~16, no.~5, pp. 14--21, 2018.

\bibitem{clouatre2022linear}
M.~Clou{\^a}tr{\'e}, M.~Balas, V.~Gehlot, and J.~Valasek, ``Linear quantum state observers,'' \emph{IEEE Transactions on Quantum Engineering}, vol.~3, pp. 1--10, 2022.

\bibitem{755839}
L.~Grover, ``Quantum computation,'' \emph{IEEE Potentials}, vol.~18, no.~2, pp. 4--8, 1999.

\bibitem{kelley1999iterative}
C.~T. Kelley, \emph{Iterative methods for optimization}.\hskip 1em plus 0.5em minus 0.4em\relax Philadelphia, PA, USA: SIAM, 1999, no.~18.

\bibitem{wyers2013bounded}
E.~J. Wyers, M.~B. Steer, C.~Kelley, and P.~D. Franzon, ``A bounded and discretized nelder-mead algorithm suitable for rfic calibration,'' \emph{IEEE Transactions on Circuits and Systems I: Regular Papers}, vol.~60, no.~7, pp. 1787--1799, 2013.

\bibitem{nakahira2016finding}
K.~Nakahira, T.~S. Usuda, and K.~Kato, ``Finding optimal solutions for generalized quantum state discrimination problems,'' \emph{IEEE Transactions on Information Theory}, vol.~63, no.~12, pp. 7845--7856, 2016.

\end{thebibliography}

\vfill

\end{document}